\documentclass[aps,prl,reprint,groupedaddress]{revtex4-1}

\usepackage{lineno,graphicx, amssymb}
\usepackage{amsmath}
\usepackage{color}
\usepackage{float}

\definecolor{purple}{rgb}{0.6, 0.2, 0.8}

\usepackage{amsmath}
\DeclareMathOperator{\sign}{sgn}

\begin{document}

\setkeys{Gin}{draft=false}


\title{Machine-learning based discovery of missing physical processes in  radiation belt  modeling}



\author{Enrico Camporeale}
\email{enrico.camporeale@noaa.gov}
\affiliation{CIRES, University of Colorado \& NOAA Space Weather Prediction Center, Boulder, CO, USA}

\author{George J. Wilkie}\affiliation{Princeton Plasma Physics Laboratory, Princeton, NJ, USA}
\author{Alexander Drozdov}\affiliation{University of California Los Angeles, CA, USA}
\author{Jacob Bortnik}\affiliation{University of California Los Angeles, CA, USA}



\date{\today}

\begin{abstract}
Real-time prediction of the dynamics of energetic electrons in Earth's radiation belts incorporating incomplete observation data is important to protect valuable artificial satellites and to understand their physical processes. Traditionally, reduced models have employed a diffusion equation based on the quasilinear approximation. Using a Physics-Informed Neural Network (PINN) framework, we train and test a model based on Van Allen Probe data. We present a recipe for gleaning physical insight from solving the ill-posed inverse problem of inferring model coefficients from data using PINNs. With this, it is discovered that the dynamics of ``killer electrons'' is described more accurately instead by a drift-diffusion equation. A parameterization for the diffusion and drift coefficients, which is both simpler and more accurate than existing models, is presented.
\end{abstract}


\maketitle

\section{Introduction}
The mechanisms that regulate the acceleration, transport, and loss of energetic particles in the Earth's radiation belts have long been investigated, both from the standpoint of fundamental research, and for practical space weather applications \cite{horne2005}. In this region, so-called ‘killer’ electrons can be accelerated to relativistic energies in just a few hours, posing a threat to satellites \cite{horne2007}.
The radiation belts consist of a collisionless, tenuous plasma, whose particles obey Maxwell's equations and whose distribution can be described by the Vlasov equation. However, due to the massive temporal and spatial scale separation of the leading physical processes, the customary approach to studying radiation belt electrons is to use a model reduction known as the quasi-linear theory, introduced in the seminal paper \cite{kennel1966}, and soon adopted in radiation belt physics \cite{lyons1972, summers1998}. 
The motion of charged particles in a dipolar magnetic field can be decomposed into three quasi-periodic orbits.
{ The first adiabatic invariant is associated with energy and pitch angle, the second invariant is associated with pitch angle, and the third invariant is associated with location and pitch angle.} In the quasi-linear procedure one can expand particle orbits around their unperturbed trajectories in the Vlasov-Maxwell equations, and derive a drift-diffusion equation in adiabatic invariant space \cite{schulz2012}. The scattering due to resonant wave-particle interactions violates the conservation of adiabatic invariants and it is responsible for most of the particle dynamics (since collisions are absent in this tenuous plasma environment). These effects can be described by the the drift and diffusion coefficients, hence dramatically reducing the complexity of the model. Furthermore, given the different timescales associated to the three adiabatic invariants, one can decouple the diffusion in the radial direction from the one in energy and pitch angle, ending up with a one-dimensional Fokker-Planck equation, valid for particles at a constant value of the first and second adiabatic invariants. Following a standard derivation (see, e.g. \cite{chandrasekhar1943}) the one-dimensional Fokker-Planck equation is:

\begin{linenomath*}
\begin{equation}\label{eq:FP}
 \frac{\partial f(\Phi,t)}{\partial t}=\frac{1}{2}\frac{\partial^2}{\partial \Phi^2}(D_\Phi f(\Phi,t)) -\frac{\partial}{\partial \Phi}(C_\Phi f(\Phi,t))
\end{equation}
\end{linenomath*}

where $f$ is the particles' Phase Space Density (PSD), $\Phi$ is the third adiabatic invariant (magnetic flux enclosed by a drift shell), $t$ is time, and Eq.(\ref{eq:FP}) is understood to be valid for constant values of first and second adiabatic invariants. The drift and diffusion coefficients ($C_\Phi$ and $D_\Phi$, respectively) have the physical meaning of mean displacement and mean square displacement per unit time.
Typically, Eq. (\ref{eq:FP}) is further simplified by assuming a simple relationship between $C_\Phi$ and $D_\Phi$, which can be derived in the case of a dipole field \cite{falthammar1966} or in absence of source or sinks \cite{roederer2016}: $C_\Phi = 1/2 (\partial D_\Phi/\partial \Phi)$ so that, upon transforming $\Phi$ to the normalized equatorial radial distance $L$ we get the familiar expression:	

\begin{linenomath*}
\begin{equation}\label{eq:diff}
 \frac{\partial f(L,t)}{\partial t}=L^2\frac{\partial}{\partial L}\left(\frac{D_{LL}}{L^2}\frac{\partial f(L,t)}{\partial L}  \right).
\end{equation}
\end{linenomath*}

Eq.(\ref{eq:diff}) has constituted the backbone of a large part of radiation belt research for the past 60 years, and even though it is now understood that energy and pitch angle diffusion are crucial ingredients for an accurate description of electrons dynamics \cite{shprits2009,thorne2010,xiao2010}, the relative importance of radial diffusion is still vigorously debated \cite{lejosne2020}. 
Although the radial diffusion coefficient $D_{LL}$ can be calculated from first-principles \cite{liu2016}, as well as for event-specific cases \cite{tu2012,li2020} (keeping in mind the several assumptions built in the quasi-linear approximation \cite{camporeale2015}), its specification requires detailed knowledge about the power spectrum and distribution of Ultra Low Frequency (ULF) waves that are resonant with electrons \cite{ozeke2012,dimitrakoudis2015}. Hence, most of the focus has been centered on finding an efficient and accurate empirical parameterization of the diffusion coefficient, possibly as a function of quantities that are available in real-time. The parameterizations most used in the literature use the geomagnetic index $Kp$ as the main driver. The parameterization by \cite{brautigam2000} (henceforth BA)  is possibly the most widely used parameterization of $D_{LL}$ as a simple function of $Kp$ and $L$. More recent works include Refs. \cite{ozeke2014,lejosne2019,ali2016,drozdov2020,wang2020}. A Bayesian approach that accounts for possible source of uncertainties has been presented in \cite{sarma2020}.

Here, we approach the problem of defining and parameterizing the coefficients of the radial transport equation from a purely data-driven standpoint and, for the first time, using machine learning techniques.
Since Eq.(\ref{eq:diff}) does not account for any injection or loss due to non-diffusive processes, we focus on the more general drift-diffusion equation:

\begin{linenomath*}
\begin{equation}\label{eq:adv_diff}
 \frac{\partial f(L,t)}{\partial t}=L^2\frac{\partial}{\partial L}\left(\frac{D_{LL}}{L^2}\frac{\partial f(L,t)}{\partial L}  \right) - \frac{\partial Cf(L,t)}{\partial L}, 
\end{equation}
\end{linenomath*}

with $C(L,t)$ a positive-definite drift coefficient. The positiveness of $C$ imposes a constraint on the solution, yet still allowing the drift term to effectively act as both a source or a loss term (i.e., it can be either positive or negative, depending on the sign of the derivative). In other words, we seek a solution of the Fokker-Planck equation in drift-diffusion form, without assuming any relationship between the drift and diffusion coefficients, {since in general $C_\Phi \neq 1/2 (\partial D_\Phi/\partial \Phi)$}. { The additional drift term is physically related to rapid particle injections into the inner magnetosphere which have often been observed by satellites, and which are not a result of a Fick’s law type inward diffusive flow, due to gradients in the diffusion coefficient, but a rapid advective flow (see, e.g. \cite{bortnik2008nonlinear}).}

To solve this inverse problem, we use a Physics Informed Neural Network \cite{raissi2019} (PINN), that derives $f$, $D_{LL}$, and $C$ as general smooth functions of $L$ and $t$, by enforcing both consistency with data and a small residual of the drift-diffusion equation (\ref{eq:adv_diff}). We use three years of Van Allen Probes data (that we consider 'noiseless') in the inverse-problem { (the training set covers the period 01-Nov-2013 to 30-Oct-2016)}. The procedure approximates the phase space density $f$ by means of a neural network (learning from the observed data), and learns $D_{LL}$ and $C$ as the optimal coefficients that solve Eq. (\ref{eq:adv_diff}) for the approximated $f$. We emphasize that all of the physics of interest and the particle dynamics are encoded in those coefficients, whose analysis then becomes extremely insightful.

The aim of this work is to perform data-driven discovery of the physics which is missing in the traditional quasi-linear diffusion equation, routinely used to study electrons in the radiation belts.
The result is twofold. First, we show that the drift term is often comparable with the diffusion one, and we analyze in detail their relative importance, with varying $L$, geomagnetic activity, and phase space density values. Second, we derive what is possibly the simplest and most interpretable parameterization of drift and diffusion coefficients as functions of $L$ only, that is still able to capture most of the dynamics, and is indeed comparable or superior in accuracy to parameterizations published earlier.

\section{Results}
\begin{figure*}[ht]
 \includegraphics[width=1\textwidth]{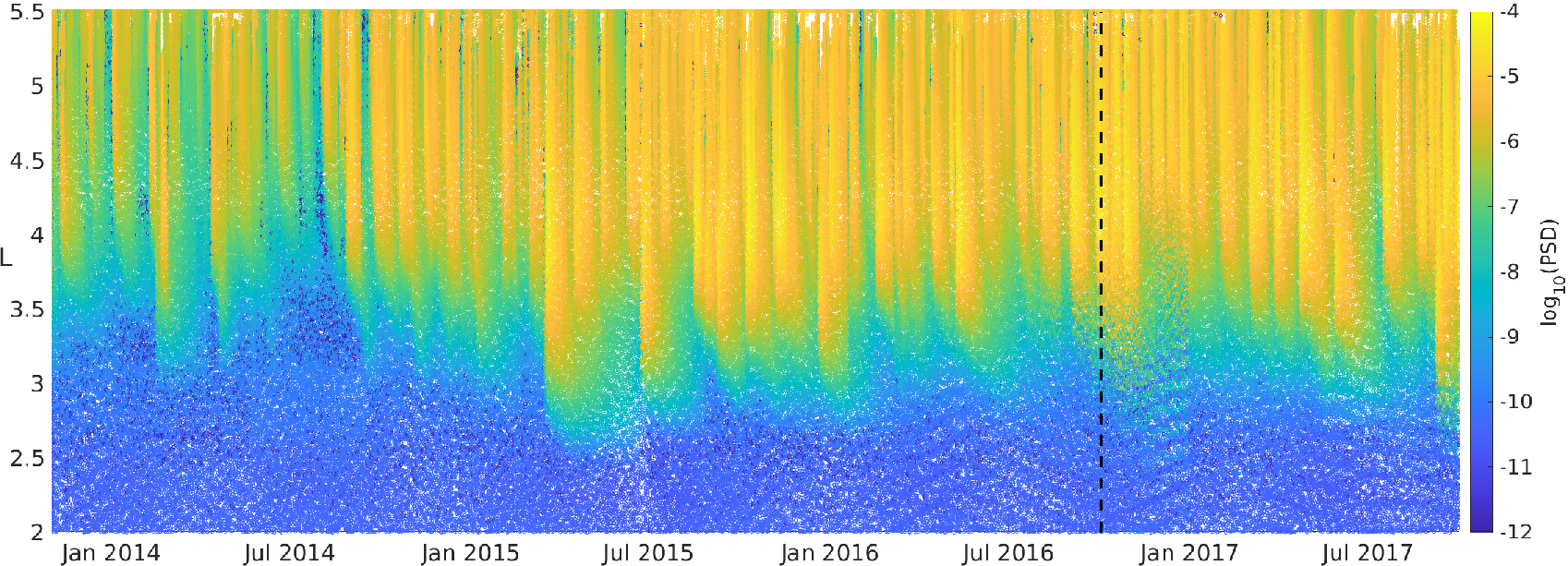}
\caption{Phase Space Density of the whole dataset, on a logarithmic scale, as function of $L$-shell. The vertical dashed line divides the dataset into a contiguous training (70\% of the dataset, to the left) and test sets (30\% of the dataset, to the right).}\label{fig:alldata}
\end{figure*}

\begin{figure*}
\centering
 \includegraphics[width=\textwidth]{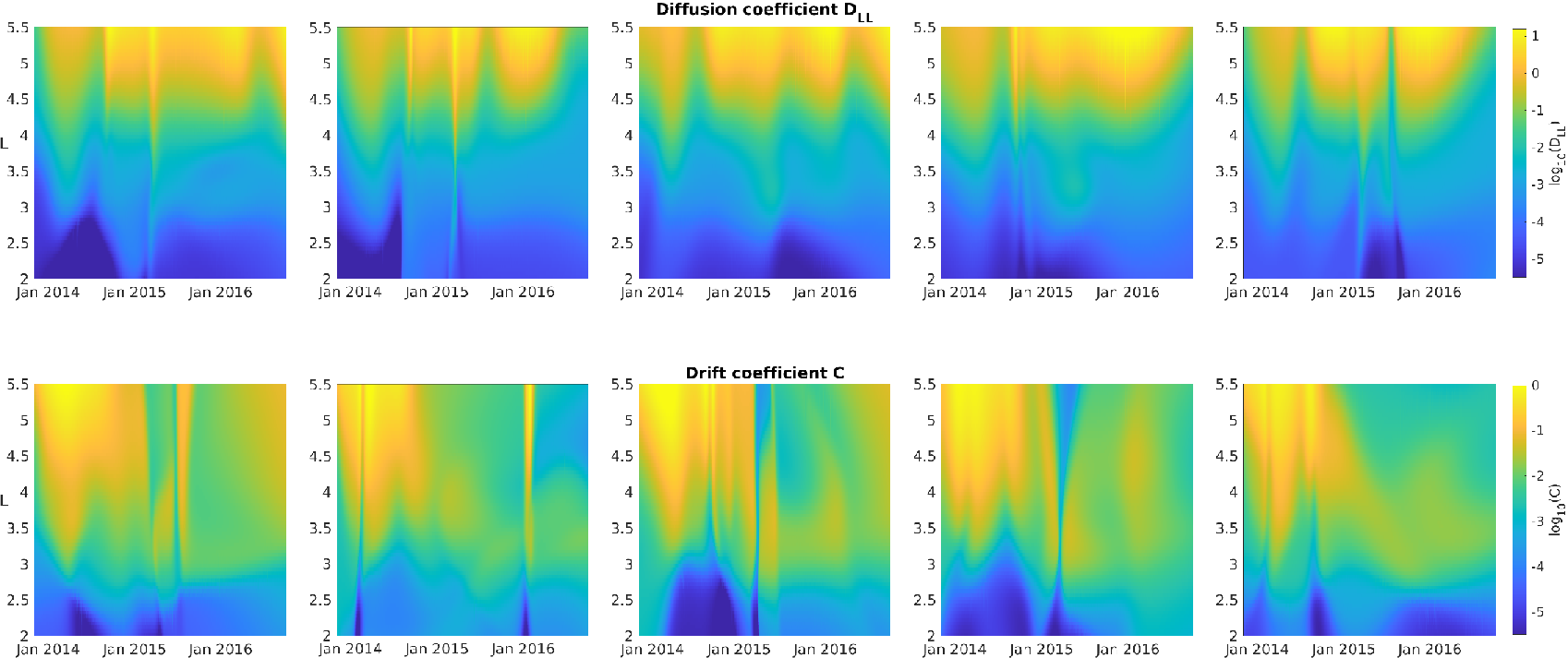}
\caption{Top 5 diffusion coefficients (top) and corresponding drift coefficients (bottom), on a logarithmic scale.}\label{fig:ens_coeffs}
\end{figure*}

\begin{figure*}[ht]
\noindent
\includegraphics[width=.8\textwidth]{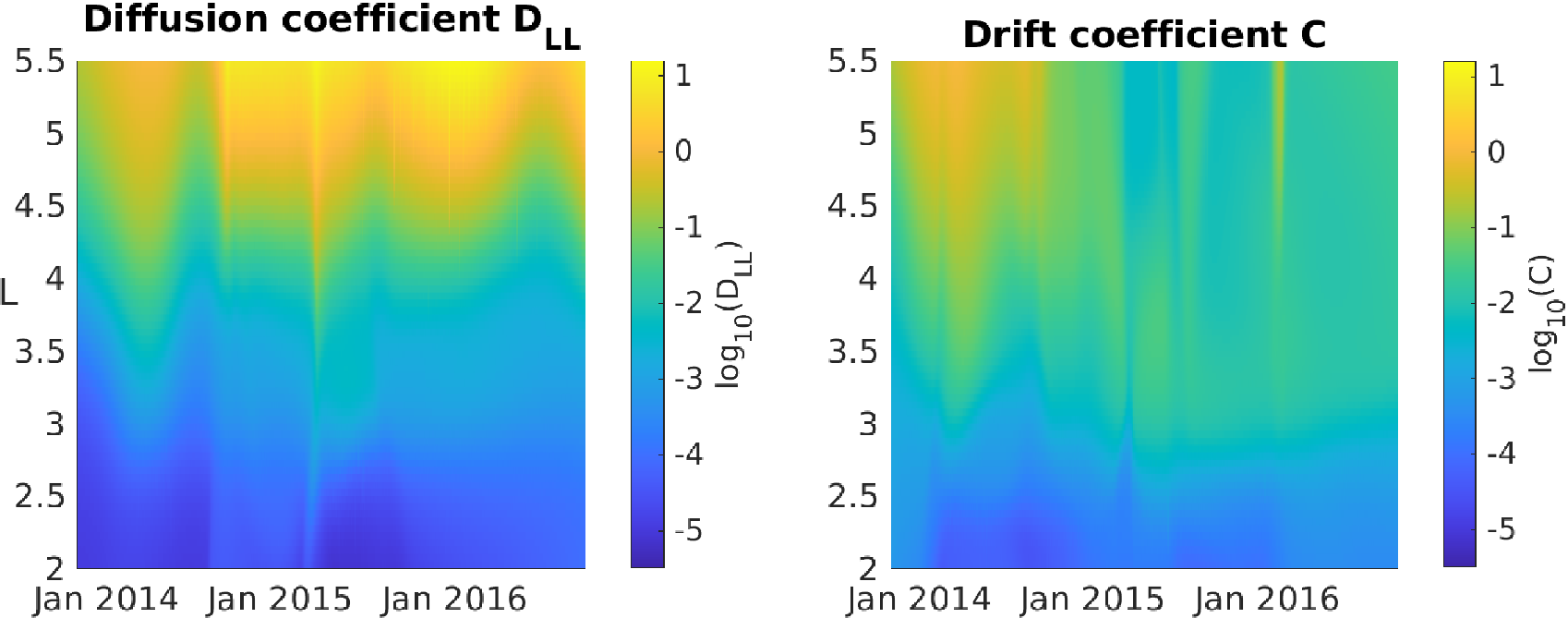}
\caption{Diffusion (left) and drift (right) coefficients obtained by averaging the top 5 solutions shown in Figure \ref{fig:ens_coeffs}}\label{fig:ens_mean}
\end{figure*}

We have applied the PINN method to solve the inverse problem of Eq. \ref{eq:adv_diff} with Van Allen Probes data used as training set (about 25,000 time instances, from 01-Nov-2013 to 30-Oct-2016, see Figure \ref{fig:alldata}).  Characteristic of solving inverse problems, the solution is not unique, so we have followed an ensemble approach, by training 20 independent PINNs. The best 5 results (in terms of smallest errors $\varepsilon$, see Methods section) are shown in Figure \ref{fig:ens_coeffs} (top and bottom panels: Diffusion coefficients and corresponding drift coefficients). We notice that, although each solution is different, they all share common large scale features in time and space. { We have also verified that the results described in the following do not substantially depend on the number of PINNs trained (i.e. the results are well converged). }
In Figure \ref{fig:ens_mean}, we show a statistical analysis applied to the \emph{optimal} diffusion and drift coefficients $D_{LL}$ and $C$, derived as an average of the 5 top solutions.

\begin{figure*}
 \includegraphics[width=.8\textwidth]{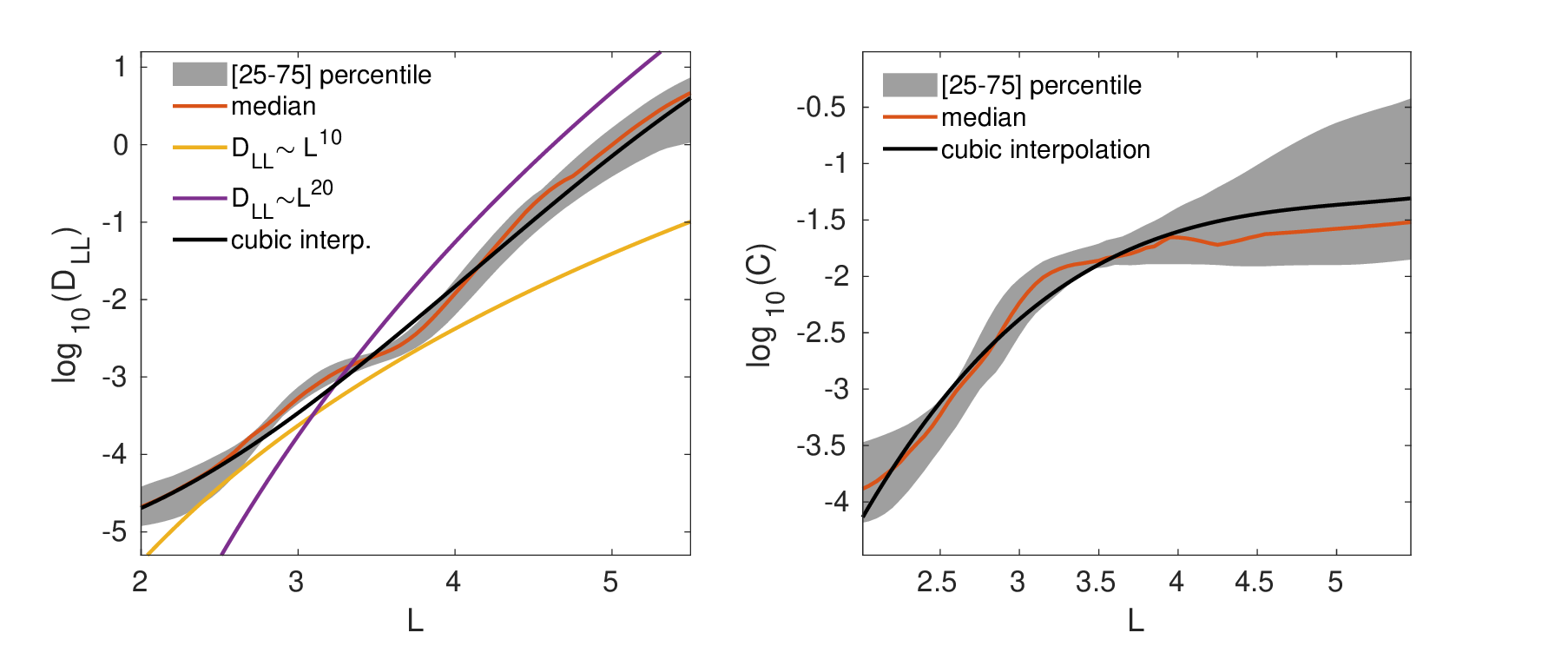}
\caption{Distribution of the diffusion coefficients $D_{LL}$ (left) and $C$ (right) as function of $L$-shell. The gray area represents the interval between the 25th and 75th percentile (for a given $L$-shell), and the orange line denotes the median. The yellow and magenta lines are shown as a reference for $L^{10}$ and $L^{20}$, respectively. The black line is a cubic interpolation fit.}\label{fig:dist_DLL_C}
\end{figure*}

Figure \ref{fig:dist_DLL_C} (left panel) shows the distribution of the diffusion coefficient $D_{LL}$ as a function of $L$. The gray area represents the interval between the 25th and 75th percentile (for a given $L$), and the orange line denotes the median. One can notice that the spread increases by moving further away from the point $L\sim 3.2$, where it reaches its minimum. Also, the slope of the distribution undergoes several regimes. For reference, we overlay the curves $L^{10}$ (yellow) and $L^{20}$ (magenta). The former is adopted in the BA parameterization \cite{brautigam2000} and is consistent with the distribution of $D_{LL}$ for small $L$, while for large $L$, the latter dependence seems more appropriate. The right panel of Figure \ref{fig:dist_DLL_C} shows the distribution for the drift coefficient $C$, with same format. One can notice two different regimes being approximately separated at $L\sim 3.5$. For $L>3.5$, $C$ can vary by one or two orders of magnitude. In both panels the black line denotes a simple cubic interpolation, described later. { The presence of (at least) two distinct regimes confirms that the physics of interest is different within and outside the plasmapause. Here we do not explicitly model the plasmapause location (see, e.g. \cite{malaspina2020wave,guo2021prediction,chu2017}), hence the change in the distributions slopes between L=3 and L=3.5 should be attributed to a statistically average plasmapause location. The spread in the coefficients is harder to interpret physically, although certainly driven by the boundary conditions at $L=2$ and $L=5.5$. We note that one of the important aspects of PINN-based insight discovery is identifying regions in parameter space that are poorly constrained or carry greater error, as specific areas that require better understanding and further investigation.}

\begin{figure}
 \includegraphics[width=.5\textwidth]{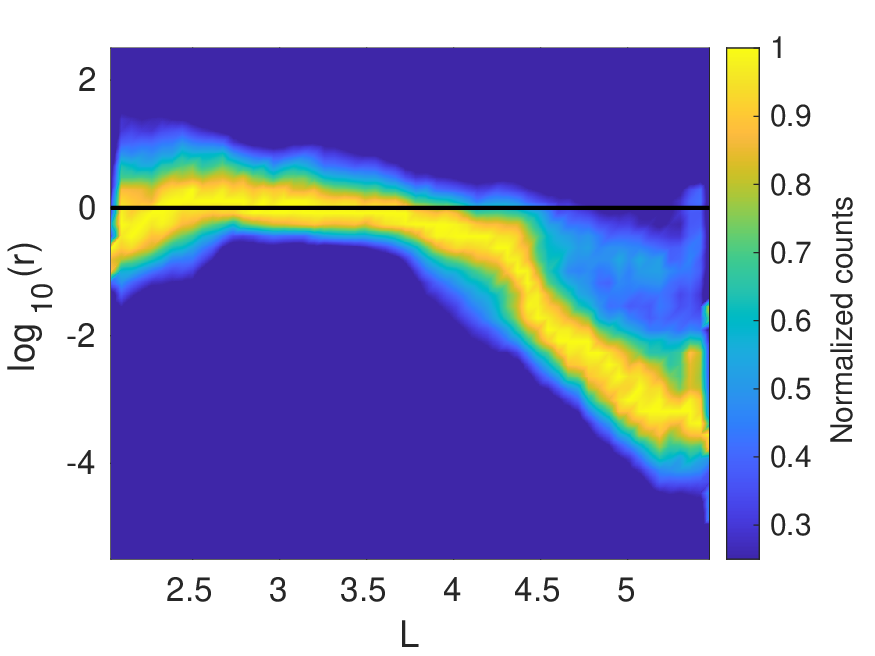}
\caption{Distribution of $r$ (logarithmic scale) as a function of $L$. The number of counts is normalized, for each value of $L$, to its maximum value. The black solid line denotes $r=1$, that is exact balance between the drift and diffusion terms.}\label{fig:dist_r}
\end{figure}

In order to understand the relative importance of the diffusion and drift terms in Eq. (\ref{eq:adv_diff}) we define their ratio as $ r= \left| \frac{1}{L^2}\left(\frac{\partial Cf}{\partial L}\right) \big/ \left[\frac{\partial}{\partial L}\left(\frac{D_{LL}}{L^2}\frac{\partial f}{\partial L}  \right)\right] \right|$. Figure \ref{fig:dist_r} shows the distribution of $r$ (in logarithmic scale, vertical axis) as a function of $L$ (horizontal axis). The distribution is normalized to the maximum value of counts per $L$-value. The black solid line at $\log_{10}r=0$ indicates equal balance between drift and diffusion, and the region below that line represents a stronger diffusion than drift. One can notice that in the inner magnetosphere ($L\lesssim 4$) the two terms are approximately balanced, while diffusion plays a larger role with increasing $L$ in the outer belt. 
{ Figure \ref{fig:dist_r} can be interpreted in the sense of local versus global losses, where the former are captured by the drift term and the latter by the diffusion term. Typically, local diffusion at $\mu=700$ MeV/G is controlled by the hiss and chorus waves and radial diffusion becomes very low at lower L-shell. On the other hand, hiss waves will more likely be a cause of local losses at low L-shell, providing a steady decay time, shorter than the one due to radial diffusion. It is important to notice that this picture might change for lower $\mu$ values, which is something that can be explored in the future using this technique.}

\begin{figure*}
 \includegraphics[width=1\textwidth]{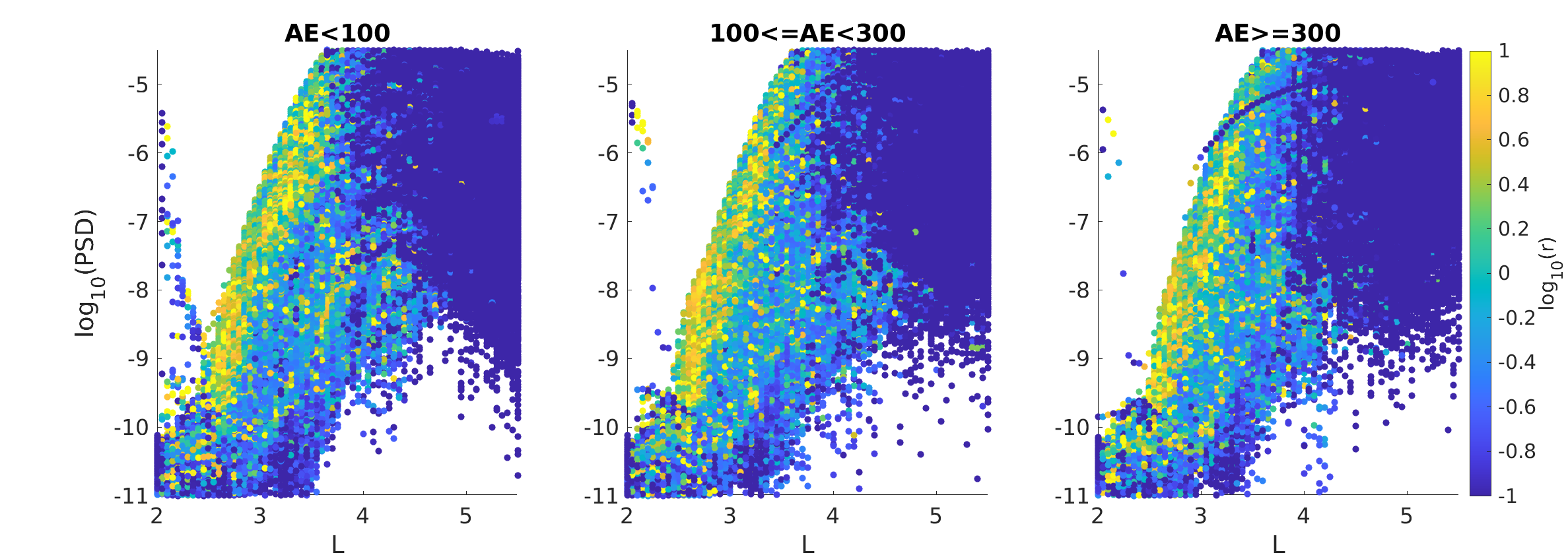}
\caption{Distribution of $r$ (logarithmic scale) as a function of $L$ and $\log_{10}(PSD)$ for three geomagnetic levels (left panel: $AE<100$, middle panel: $100\leq AE<300$, right panel: $AE>=300$)}\label{fig:scatter}
\end{figure*}

We further analyze the relative contribution of the drift and diffusion terms by studying the ratio $r$ as a function of $\log_{10} f$ and $L$, and for different geomagnetic activity, represented by the Auroral Electrojet index AE, in Figure \ref{fig:scatter} (left panel: $AE<100$, middle panel: $100\leq AE<300$, right panel: $AE>=300$). Interestingly, at low $L$ drift is more dominant than diffusion for larger values of PSD. Also, the range of $L$ in which diffusion is dominant slightly shifts to smaller $L$ with increasing geomagnetic activity. This analysis unambiguously shows an unexpected relatively large contribution of non-diffusive drift in the time evolution of the phase space density. 

Finally, we discuss how the PINN-derived drift and diffusion coefficients can be used for deriving a very simple and interpretable parameterization that can be used in forward simulations. 
A standard feature selection procedure (not shown) demonstrates that most of the variance in both $D_{LL}$ and $C$ can be attributed to changes in $L$. In other words, $L$ is the best unique predictor for the coefficients, and therefore we aim to describe them as a function of $L$ only, by fitting the PINN-derived values of $D_{LL}$ and $C$ with a cubic interpolator, shown with black lines in Figure \ref{fig:dist_DLL_C}. The derived formulas for the cubic fit are the following:
\begin{linenomath*}
\begin{eqnarray}
 \log_{10}D_{LL} = -0.0593L^3 + 0.7368L^2 -1.33L -4.505\label{eq:cubic1} \\
 \log_{10}C = 0.0777L^3 -1.2022L^2 + 6.3177 L -12.6115\label{eq:cubic2}
\end{eqnarray}
\end{linenomath*}

In order to assess the goodness of this approximation, we use it in a forward model solution (see section Methods) and we compare the results with two benchmarks: a solution derived with the BA diffusion coefficients \cite{brautigam2000}, and another derived by using the diffusion coefficients proposed in Ozeke et al. \cite{ozeke2014}. For both cases we solve Eq. (\ref{eq:diff}) with the addition of a loss term ($-f/\tau$), parameterized as in \cite{gu2012,orlova2016}, since the inclusion of such term is standard practice to account for wave-particle scattering due to hiss and chorus waves, and it is known to improve accuracy.
{ In Fig. \ref{fig:err_test_cubic} we show the percentage symmetric accuracy $\zeta$, Eq. \ref{eq:zeta} (left) and the symmetric signed percentage bias SSPB, Eq. \ref{eq:SSPB} (right) (see Methods) calculated over the whole test set (1 year of data), as a function of $L$. 
Blue, red, and black lines denote the results from the baselines by BA and Ozeke et al., and by using the PINN-derived cubic fit, respectively. In the left panel of Fig. \ref{fig:err_test_cubic}, the solid squares denote the median values $\zeta_{50}$ and the error bars are calculated as the spread between $\zeta_{25}$ and $\zeta_{75}$. In the right panel, positive values are in solid and negative values in dashed lines.
One can notice that the simple cubic approximation of Eqs. (\ref{eq:cubic1},\ref{eq:cubic2}) is comparable or superior to the results obtained with more sophisticated models. Furthermore, one can notice that all errors are by definition going to zero at the boundary, and in fact the boundary conditions are possibly the most important driver of the dynamics at large $L$.}\\

{ Finally, we present in Figure \ref{fig:err_test_cut} the PSD resulting from the forward models using the three different parameterizations (BA in red, Ozeke et al. in yellow and PINN-derived cubic fit in purple), compared against the Van Allen Probes data (blue), for the whole period cover in the test set. Top and bottom panels are for $L=5$ and $L=4$, respectively. In all cases, the simulations have initial and boundary conditions taken from the data. For $L=5$, the PSD resulting from the new parameterization presented here is consistently more accurate than the two baseline models, which tend to underestimate the Phase Space Density. At $L=4$ none of the three models is particularly accurate, although the PINN is often orders of magnitude closer to the observations than the other two models. Note that logarithmic scales are used in vertical axis.} 

\section{Discussion}
The process of understanding the mechanisms underlying a physical process, and the ability of describing such mechanisms with the elegant and succinct formalism of partial differential equations (PDEs) lies at the core of scientific discovery. However, the way in which a scientists extracts information from experiments and observations (\emph{data}) and encodes that information into PDEs has seen dramatic changes over the last decade, when methods originating in machine learning have started playing an increasingly important role. Currently, there is a rich literature on data-driven discovery of PDEs (see, e.g., \cite{long2018, berg2019,raissi2018deep,rudy2017,xu2019,zhang2018,boulle2021,udrescu2020}). The published methods can be loosely divided in two classes. On one hand, one can create a large dictionary of terms that contain algebraic, differential and integral operators and search the space of all (or many) combinations of those terms for the optimal PDE that describes the data (i.e., the PDE whose solution is an acceptable approximation of the data). Two seminal examples of this approach are Ref. \cite{rudy2017} (using sparse regression) and Ref. \cite{udrescu2020} (using symbolic regression). On the other hand, one can restrict the search for the optimal PDE to a specific class of functionals, thus setting up the problem of PDE discovery as an inverse problem, where the time and space dependence of free parameters (such as, for instance, drift and diffusion coefficients) needs to be learned. Physics-Informed Neural Network, introduced in \cite{raissi2019}, falls in this category, and it is the approach used in this paper.

\begin{figure*}
\centering
\includegraphics[width=\textwidth]{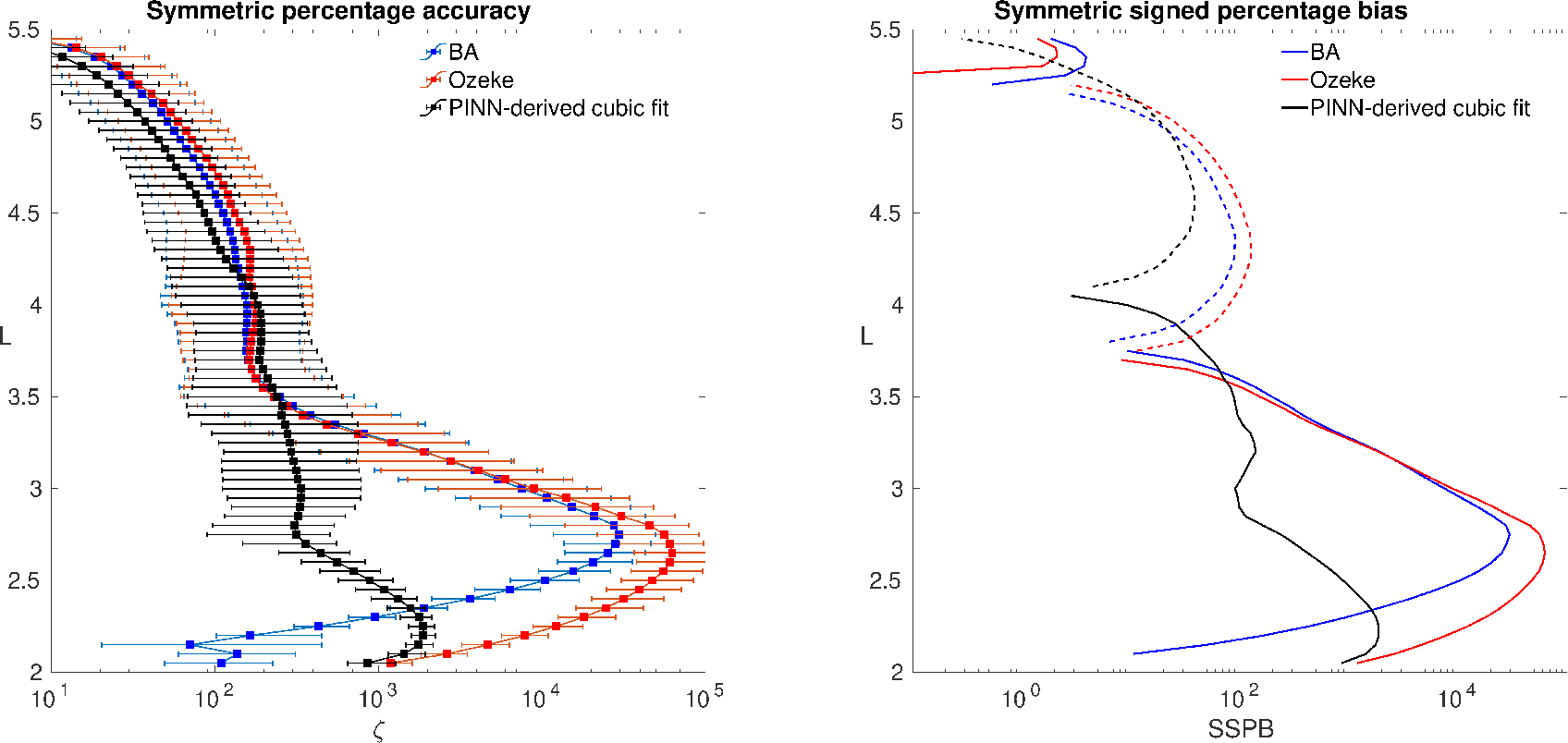}
\caption{{ Percentage symmetric accuracy $\zeta$ (Eq. \ref{eq:zeta}) (left) and symmetric signed percentage bias SSPB (Eq. \ref{eq:SSPB})(right) calculated over the whole test set (1 year of data), as a function of $L$. Blue and red lines denotes the BA and Ozeke et al. baseline models, respectively, while the cubic parameterization in Eqs. (\ref{eq:cubic1}-\ref{eq:cubic2}) is shown in black. In the left panel, the solid squares denote the median values $\zeta_{50}$ and the error bars are calculated as the spread between $\zeta_{25}$ and $\zeta_{75}$. In the right panel, positive values are in solid and negative values in dashed lines.}}\label{fig:err_test_cubic}
\end{figure*}

\begin{figure*}
\centering
\includegraphics[width=\textwidth]{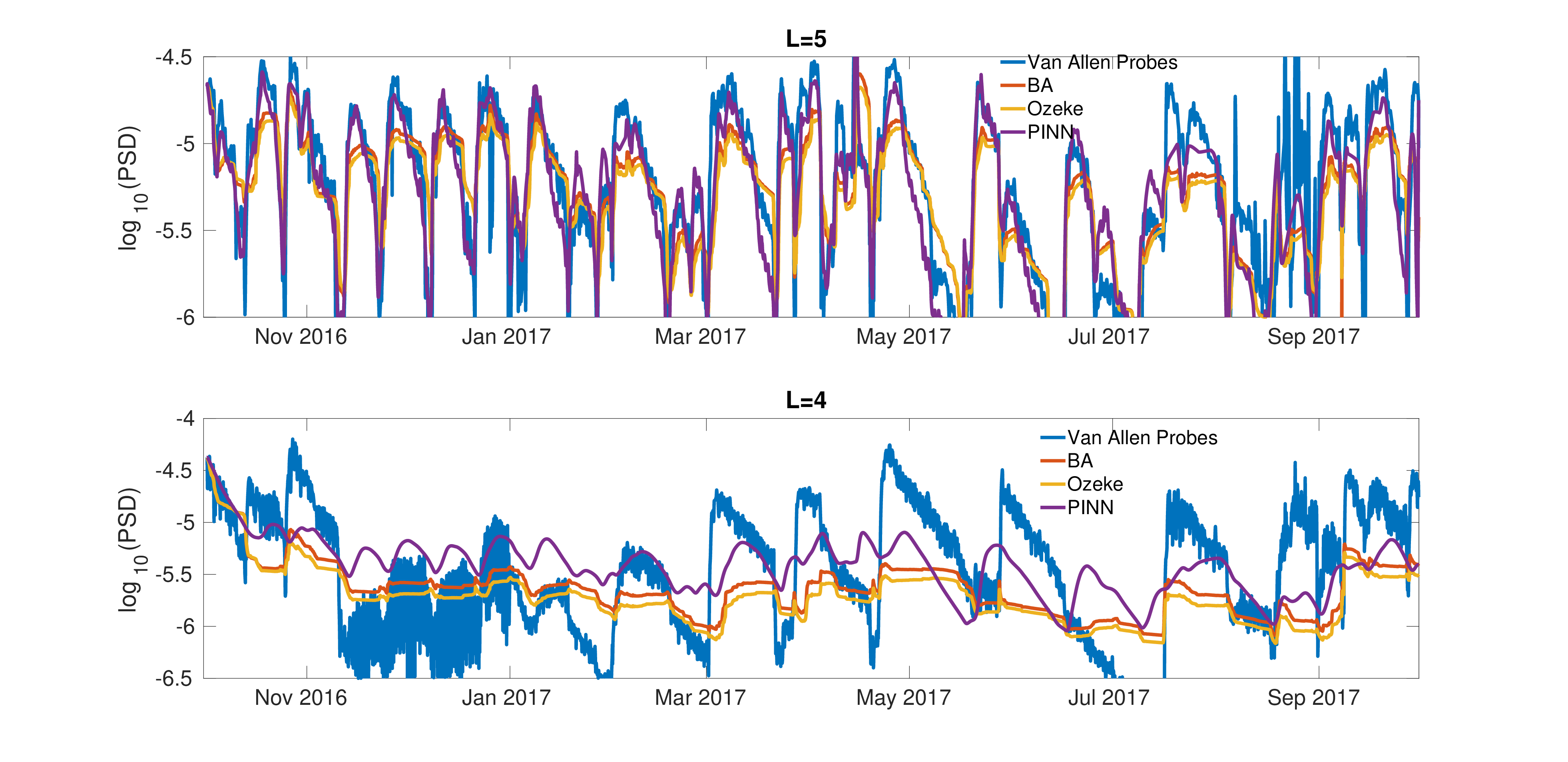}
\caption{{ Phase Space Density (PSD) resulting from running the forward model with different coefficient parameterization, for the while test set. Red, yellow and purple lines denote the BA, Ozeke et al. and PINN-derived cubic parameterizations, respectively. The Van Allen Probes data is represented in blue. The vertical axis is in logarithmic scale.}}\label{fig:err_test_cut}
\end{figure*}

Specifically, we have investigated here the possibility that the time evolution of the Phase Space Density of electrons in the Earth's radiation belt could be described by the combination of (and the competition between) a diffusion and a drift term. On one hand, the hypothesis that a non-diffusive drift mechanism might take place is in line with the general procedure of deriving a Fokker-Planck equation from the first-principles Vlasov equation. On the other hand, this hypothesis challenges several decades of literature that have exclusively focused on diffusive processes (more or less implicitly justified by assuming a relationship between drift and diffusion coefficients of the form discussed in the Introduction).
The data-driven approach enabled by PINN allows to unambiguously test such hypothesis, by determining the optimal drift and diffusion coefficients that, used in Eq. (\ref{eq:adv_diff}), result in the solution most consistent with observations. Obviously, as powerful as it is, the PINN method does not solve the issue of ill-posedness of the inverse problem. Namely, there is no guarantee about the uniqueness of the solution. Indeed, we have verified that different realizations of the coefficients are possible and equally valid. Interestingly enough, we have also verified that not only the best 5 coefficients used in this study yield solutions that have comparable errors with respect to the data, but that the average of the coefficients (analyzed in detail in Figures \ref{fig:ens_mean}-\ref{fig:scatter}) also yield a similar level of error.\\
Finally, in the grand scheme of scientific machine learning \cite{roscher2020explainable}, one would like to use advanced but often opaque techniques (such as PINN) to extract physical insight from the data, but with the final goal of exploiting such new insights to eventually advance our knowledge and possibly derive new interpretable models. In a sense, such grand scheme follows the old argument of Occam's razor that suggests that one should seek the most parsimonious yet accurate model. 

In this spirit, we have used the PINN-discovered coefficients $D_{LL}$ and $C$ and their learned dependence on $L$ to build a simple and interpretable model (with no free parameters, other than the boundary conditions) that yields an excellent approximation (and forecast) of the PSD (Figures \ref{fig:err_test_cubic},\ref{fig:err_test_cut}) when compared to the state-of-the-art models (that, instead, exploit future knowledge of geomagnetic activity in the form of $Kp$). In our opinion, this last step represents the pinnacle of scientific machine learning, where a simple, analytical, interpretable expression for physical parameters has been discovered by way of using a powerful, yet opaque, ML method such as PINN.\\

In conclusion, we have used the formalism of PINN to solve the inverse problem of a drift-diffusion equation for energetic electrons in the radiation belt to evaluate the importance of a { non-diffusive} drift mechanism { in L} that has so far been overlooked in the literature. We have discovered that the drift term is non-negligible and often comparable with the diffusion one, and we have studied their relative importance as a function of $L$, geomagnetic activity, and phase space density values. Finally, we have derived a parameterization of the drift and diffusion coefficients as functions of $L$ only, which turns out to be their best unique predictor. A simple cubic fit has been tested on held-out data (test set). The accuracy of this new PINN-derived parameterization is comparable with and often out-performs baseline models, routinely used in the literature. Because the new parameterization does not depend on $Kp$, it can be straightforwardly implemented in space weather forecasting.


\section{Methods}
\subsection{Data}
We use observations from the Magnetic Electron Ion Spectrometer (MagEIS) instruments aboard the Van Allen Probes spacecraft \cite{blake2013}.
Van Allen Probes is a NASA twin satellite mission that was active for 7 years, since its launch on August 30th, 2012. Its primary mission was to address how populations of high energy charged particles are created, lost and dynamically evolve within Earth's magnetic trapping region \cite{fox2014}. Due to the unprecedented quality and quantity of data collected, Van Allen Probes have marked a golden era for radiation belt studies \cite{li2019}.
Here, we limit our study to electrons with first adiabatic invariant $\mu=700$ MeV/G and second adiabatic invariant $K=0.1$ $R_E$ G$^{0.5}$, which corresponds to approximately 1 MeV electron energies in the heart of radiation belt and are near-equatorially trapped. We used TS05 magnetic field model \cite{Tsyganenko2005-vz} to calculate the adiabatic invariants.
The dataset is comprised of $\sim$570,000 data points spanning the time range 01-Nov-2013 to 30-Sep-2017. The largest interval between consecutive data points is 2:45 hours, and the average interval is about 4.5 minutes.

\subsection{Forward model}
Eq.(\ref{eq:adv_diff}) is solved by means of an unconditionally stable, second order accurate, Crank-Nicholson scheme discussed in \cite{welling2012}. For completeness, we report the numerical discretization here:

\begin{linenomath*}
\begin{multline}\label{eq:discrete}
 \frac{f^{n+1}_j - f^n_j}{\Delta t} = \frac{L^2_j}{2\Delta L^2}\left[D^{n+\frac{1}{2}}_{j+\frac{1}{2}}(f_{j+1}^n-f^n_j + f_{j+1}^{n+1} - f_j^{n+1})\right. \\ \left. - D^{n+\frac{1}{2}}_{j-\frac{1}{2}}(f_{j}^n-f^n_{j-1}+f_j^{n+1} - f^{n+1}_{j-1}) \right] \\
 -\frac{1}{4\Delta L}\left[C^{n+\frac{1}{2}}_{j+1}(f^{n+1}_{j+1}+f^{n}_{j+1})- C^{n+\frac{1}{2}}_{j-1}(f^{n+1}_{j-1}+f^{n}_{j-1}) \right]
\end{multline}
\end{linenomath*}

where indexes $n$ and $j$ represent discretization in time and space, with time steps $\Delta t$ and $\Delta L$, and $D_j=D_{LL,j}/L^2_j$, respectively. Eq. (\ref{eq:discrete}) is a linear equation that can be written in matrix form with tri-diagonal matrices and is solved by a standard LU decomposition. For all the results presented, we use $\Delta t = 1$ (hours) and $\Delta L=0.05$. Observations at $L=2.0$ and $L=5.5$ are used as time-dependent boundary conditions, while initial conditions are interpolated from the data.

\subsection{Physics-Informed Neural Networks}\label{sec:PINN}
Physics-informed Neural Networks (PINN) are a framework for solving forward and inverse problems involving nonlinear partial differential equations \cite{raissi2019}. The theoretical foundation of PINNs lies on the well-known universal approximation property of neural networks \cite{hornik1989} that essentially allows neural networks to accurately approximate a large class of continuous functions. The basic idea of PINNs is rather simple, and it exploits the fact that the output of a neural network is a continuous and differentiable function (almost everywhere). Moreover, PINNs take advantage of the ability of modern neural network libraries to automatically calculate exact derivatives with respect to the input variables, by applying the chain rule of differentiation (this is known as \emph{autodiff} in machine learning jargon \cite{geron2019}). Hence, each term in a partial differential equation (PDE) can be calculated exactly on a set of collocation points within the domain, and the PDE itself can be used as penalization term in the loss function minimized by the neural network. Upon convergence, a PINN outputs a function that approximately solves the PDE and matches the given data on the points where it has been trained.\\
Because the solution $f$ spans several orders of magnitude in the $L$ domain, we perform the transformation $f=e^g$ and solve for $g$:

\begin{linenomath*}
\begin{equation}\label{eq:g}
 \frac{\partial g}{\partial t}  = L^2\frac{\partial}{\partial L}\left(\frac{D_{LL}}{L^2}\frac{\partial g}{\partial L}\right) + D_{LL}\left(\frac{\partial g}{\partial L}\right)^2  - g\frac{\partial C}{\partial L} - C \frac{\partial g}{\partial L}
\end{equation}
\end{linenomath*}

The PINN is designed as a combination of three coupled neural networks, each taking a point in $(L,t)$ as input and outputting the value of $f$, $D_{LL}$, and $C$ at that point, respectively. Those three outputs are then combined in the loss function, which is the sum of the mean square error with respect to the observations, and the residual of Eq. (\ref{eq:g}).
Boundary conditions (at $L=2$ and $L=5.5$) are enforced by neglecting the residual term in the loss function on those points (that is, the function $f$ is forced to converge to the boundary values).
{ The neural network architectures are standard, and have been selected by progressively increasing their complexity until a plateau in the loss function was observed. Other hyper-parameters were not optimized. The networks use a tanh activation function in all the layers. The network that outputs the solution $f$ uses 6 inner layers with $[30, 20, 20, 20, 20, 20]$ neurons, while the two networks outputting the coefficients $D_{LL}$ and $C$ have 3 inner layers with $[30, 20, 10]$ layers.} 
To perform the optimization we use a combination of the Adam optimizer \cite{kingma2014} and the BFGS (Broyden-Fletcher-Goldfarb-Shanno) method \cite{zhu1997}, both within the Tensorflow framework \cite{abadi2016}.

\subsection{Metrics and benchmarks}\label{sec:metrics}
Our quantity of interest, the phase space density $f$, changes by several orders of magnitude between $L=2$ and $L=5.5$. Hence, it is not straightforward to design a single metric for model performance. A through analysis of several metrics often used in radiation belt modeling, can be found in Refs. \cite{morley2018, liemohn2021}. Here, we are interested in studying the model accuracy at given values of $L$, rather than averaged over the whole domain.{ We define and use three different errors. Following Ref. \cite{morley2018}, we characterize accuracy by defining the \emph{percentage symmetric accuracy} $\zeta$ as:

\begin{linenomath*}
\begin{equation}\label{eq:zeta}
  \zeta_k = 100\cdot\exp(P_{k}(|\log (f/\hat{f})|)), 
 \end{equation}
 \end{linenomath*}

 where $\hat{f}$ and $f$ are the the ground-truth values taken by observations and the corresponding values produced by a model, respectively. $P_{k}$ represents the $k-$th percentile (i.e. $P_{50}$ is the median) calculated over all values at fixed $L$. This represents a generalization of the median symmetric accuracy \cite{morley2016alternatives} for quantiles other than the median, that allows to estimates error bars (that is, $\zeta_k$ is monotonically increasing with increasing $k$, see Figure \ref{fig:err_test_cubic}). The second metric we employ characterizes bias and is called the \emph{symmetric signed percentage bias} SSPB, again generalized from the definition in \cite{morley2018}:

 \begin{linenomath*}
\begin{equation}\label{eq:SSPB}
 {\rm{SSPB}} = 100\cdot\sign(P_{50}(\log (f/\hat{f})))(\exp(|P_{50}(\log(f/\hat{f}))|)-1)
\end{equation}
\end{linenomath*}

Note that, by taking the absolute value after calculating the percentile, SSPB is not ordered when considering different percentiles $P_k$ (hence it does not allow to estimate error bars).
Finally, we define the relative error $\varepsilon$ as the median value at fixed $L$ of the relative error of the logarithmic phase space density. That is:

\begin{linenomath*}
\begin{equation}\label{eq:error}
\varepsilon(L) = P_{50}\left(\frac{\log_{10}f - \log_{10}\hat{f}}{\log_{10}\hat{f}}\right) 
\end{equation}
\end{linenomath*}

}
We benchmark our results against two parameterizations for the diffusion coefficient: the BA model \cite{brautigam2000}, and the \citet{ozeke2014}, which are functions of $L$ and the geomagnetic index $Kp$ only \cite{rostoker1972}. Their formula are:

\begin{linenomath*}
\begin{align*}\label{eq:DLL}
  D_{LL}^{BA} &= L^{10}\cdot10^{(0.506Kp-9.325)}\\
  D_{LL}^{Ozeke} &= 2.6\cdot L^6\cdot 10^{(0.217L+0.461Kp-8)}\\ 
  &+ 6.62\cdot L^8\cdot 10^{(-0.0327L^2+0.625L-0.0108Kp^2+0.499Kp -13)}\\
\end{align*}
\end{linenomath*}

The electron lifetime is parameterized as in \cite{gu2012,orlova2016}, and the plasmapause location is calculated using the model presented in Ref. \cite{carpenter1992}.

\newpage

\section{Competing interests}
The authors declare no competing interests.
\bibliography{biblio}

\end{document}